\documentclass[a4paper,11pt,oneside]{article}
\usepackage[a4paper,left=2cm,right=2cm,top=2.5cm,bottom=3cm]{geometry}
\usepackage{amssymb, mathrsfs, amsthm}
\usepackage[centertags]{amsmath}
\usepackage{endnotes}
\usepackage{graphicx}
\usepackage{csquotes}
\usepackage{bbm}
\usepackage{subfigure}
\usepackage{mwe}
\usepackage[noblocks,affil-sl]{authblk}             
\usepackage{caption}
\usepackage{chngcntr}
\usepackage{epstopdf, color}
\usepackage{adjustbox}
\counterwithin{figure}{section}
\counterwithin{table}{section}

\theoremstyle{plain}
\newtheorem{thm}{Theorem}[section]

\newtheorem{remark}[thm]{Remark}

\newcommand\de{\mathrm{d}}

\newcommand\media{\mathbb{E}}
\newcommand\cppi{V^{CPPI}_t} 
\newcommand\obpi{V^{OBPI}_t} 
\newcommand\cppiz{V^{CPPI}_0}
\newcommand\obpiz{V^{OBPI}_0}
\newcommand\obpit{V^{OBPI}_T}

\newcommand\OO{\mathcal{O}}
\newcommand\QQ{\mathbb{Q}}
\newcommand\PP{\mathbb{P}}
\newcommand\FF{\mathbb{F}}

\numberwithin{equation}{section}

\makeatother
\begin{document}

\author[1]{L. Di Persio \thanks{luca.dipersio@univr.it}} 
\author[2]{I. Oliva \thanks{i.oliva@univpm.it (Corresponding author)}} 
\author[3]{K. Wallbaum \thanks{Kai.Wallbaum@risklab.com}} 

\affil[1]{Department of Computer Sciences, University of Verona, Italy}
\affil[2]{Department of Management, Universit\`a Politecnica delle Marche, Italy}
\affil[3]{risklab GmbH, Munich, Germany}

\title{Options on CPPI with guaranteed minimum equity exposure}


\maketitle 

\begin{abstract}
In the present paper we provide a two-step principal protection strategy obtained by combining a modification of the \emph{Constant Proportion Portfolio Insurance} (CPPI) algorithm and a classical \emph{Option Based Portfolio Insurance} (OBPI) mechanism. Such a novel approach consists in assuming that the percentage of wealth invested in stocks cannot go under a fixed level,  called \emph{guaranteed minimum equity exposure,} and using such an adjusted CPPI portfolio as the underlying of an option. The first stage ensures to overcome the so called \emph{cash-in} risk, typically related to a standard CPPI technique, while the second one guarantees the equity market participation. To show the effectiveness of our proposal we provide a detailed computational analysis within the Heston-Vasicek framework, numerically comparing the evaluation of the price of European plain vanilla options when the underlying is either a purely risky asset, a standard CPPI portfolio and a CPPI with guaranteed minimum equity exposure. 
\end{abstract} 




%
%
%

\section{Introduction} \label{Introduction}
During recent years, financial markets have been mainly characterized by the consequences of the great financial crisis happened in 2008 and, afterward, by the linked increasing of equity markets and related decreasing of interest rate levels around the world, until a significant market drop in 2018. 
While interest rate levels still remained on record lows, volatility in the market increased again significantly in the last 12 months.  

Such a market framework costitutes a big challenge for institutional as well as for retail investors, who are looking for an \emph{equity market 
participation} plus a \emph{downside protection.} 

The literature has tried to solve this puzzle in terms of \emph{portfolio insurance strategies,} at least starting from the early 70s of the last century, see, e.g., \cite{BalderBrandlMahayni09, BlPer92, BrennanSchwartz76, GrossmanZhou93}, and references therein. 
Roughly speaking, a portfolio insurance strategy is a protection blueprint, based on the definition of a fixed threshold such that the terminal portfolio value always lies above it. 
This approach bypasses 
the risk of the actual return being below the expected return, or the uncertainty about the magnitude of that difference, 
see, e.g., \cite{Hor05,McnRudEmb05}. 

The portfolio insurance strategies were first introduced in  \cite{LelRub76}, after the collapse of stock markets (the \emph{New York Stock Exchange's Dow Jones Industrial Average} and the \emph{London Stock Exchange's FT 30}, see, e.g., \cite{Davis03}) which implied the pension funds withdrawal. 
In particular, the authors noted ex-post that the presence of an insurance of the above mentioned type of risk could have convinced the investors not to leave the market, guarantying them  later the opportunity to take advantage of the rise of the same, an event that really happened just a couple of years later. In this context, the insurance can be interpreted as a put option on the whole portfolio. 

Portfolio insurance strategies 
can be pigeonholed into three different classes, see, e.g., \cite{PerShar95,Rudolf95}: the \emph{option-based strategies,} the \emph{option-duplicating strategies} and the \emph{derivative-independent strategies.} 
A technical analysis, also in terms of performance evaluation, of such strategies can be found in \cite{CesCrem03}, where bull, bear and no-trend markets are considered, while an overview of portfolio insurance strategies, along with their possible connections with financial instability, is provided in \cite{PainRand08}, and interesting  numerical insights have been reported in \cite{KosEtAl15}. 

An early approach, related to the first class of strategies, is the so called \emph{Option-Based Portfolio Insurance} (OBPI) method, see, 
e.g., \cite{BerPrig05}, which consists of buying a zero-coupon bond with maturity equal to the investment time horizon plus an option written on the portfolio risky asset. As an alternative, in \cite{ABW00} a \emph{minimum-cost portfolio insurance} strategy is presented. 
Here, the idea is to solve a portfolio optimization problem in incomplete markets by minimizing costs of a portfolio under the constraint 
that the payoff is greater than the insured one, avoiding losses and capturing gains. 

The option-duplicating strategy is an approach in which the option is replicated with a self-financing strategy, in order to overcome the lack of liquid options for long maturities. 

It is worth to mention that low interest rate levels are reducing the available risk budgets significantly, forcing practitioners to rethink at how to optimize portfolios' design by offering a sustainable equity market participation plus a capital protection of the initial investment. In this direction, one choice consists in considering dynamic risk management tools to protect portions of the initial investment by dynamically allocating both  in risky and riskless assets, based on available portfolio risk budgets. 
In this framework the \emph{Constant Portfolio Protection Insurance} 
(CPPI) is one of the most used approaches, see, e.g., \cite{AmeurPrigent11,Basak02,BlJo92,BlPer92,Jessen14,PauLac11,Schied14}, and references therein. 
The CPPI method is obtained by rebalancing an initial portfolio at each observation time, evaluating a present value of the aspired capital protection and then investing the available risk budget times a market-depending multiplier into risky assets, while investing the remaining part of the portfolio in time-congruent risk-free assets. An interesting analysis of portfolio insurance strategies, including the CPPI methodology, is provided in \cite{AOV09}, where the authors exploit Value-at-Risk, Expected Shortfall and stochastic dominance to measure the portfolio performance of the above-mentioned techniques.  

By assuming that the CPPI portfolio evolves according to a Markov process, in \cite{PauLac11} the authors focus on a discrete-time CPPI-based portfolio allocation method. More recently, a machine learning approach to determine the value of the parameters used to evaluate the correct proportion of wealth to be invested in stock is given in \cite{DehEsf18}.  

Despite a significant simplicity and a remarkable ease of implementation, the CPPI strategy suffers a fundamental drawback represented by the risk that, after a severe market draw-down, the risk budget erases with a consequent reduction to zero of the market participation, hence potentially allowing for what practitioners call the \emph{cash-in event.} To overcome the latter scenario, most of the practitioners resort to different routines, such as using an intertemporal risk budgeting which allows the full use of the available risk budget over time. Alternatively, they adopt the multiplier related to market volatility. 
However, while both methods can reduce the probability of \emph{cash-in event} to happen, they cannot guarantee to  avoid it within the traditional CPPI approaches. 

To close this gap, we introduce an innovative method consisting in a combination of a CPPI strategy and 
an OBPI one. In particular, we overcome two different problems by annihilating the cash-in risk, simultaneously ensuring some participation in upside markets. 
Our solution is based on the following: 
starting from the CPPI portfolio dynamics, a threshold 
is included in the proportion of wealth invested in stocks. Such a threshold is dubbed \emph{guaranteed minimum equity exposure} 
(GMEE) and represents the minimum investment value in the risky asset. The portfolio obtained through the previous mechanism can be used 
as underlying of an option. 

We would like to highlight that, although such an approach has been selectively used in practice, no rigorous mathematical 
treatment of it has been provided up to now, at least to the best of our knowledge. Therefore, our work represents the first rigorous 
analytical treatment toward this direction.
Even if within well establlished literature the aforementioned topics have been already described, such analysis have been separately provided.
In particular, the analysis of factors that can potentially lead to gap risk in portfolio insurance framework, mainly taking into account the asset price behavior and the trading frequency, is described in \cite{ConstEtAl08}. 
Conversely, we refer to \cite{EKZ11}, resp. to \cite{AlbStebWall17}, for related studies on options on a standard CPPI logic, resp. on options linked to the so called \emph{VolTarget} strategies. 

In this sense, the present paper can be considered the first attempt to combine the above mentioned topics in a unified framework. 
To  better explain the concreteness as well as  the goodness of our approach, we shall show how our 
proposal would have worked in the past, providing some historical simulations of structured products with CPPI and CPPI-GMEE. 
We scrutinize both versions of the CPPI logic in different market scenarios, 
to better appreciate the sensitivities of the strategies. Such an investigation enables to figure out the behaviour of the 
risk--return profile, as well as of the asset allocation in different market cycles. 
Thereafter, we also determine the prices of the corresponding CPPI options for different set of product 
levels for a market model where both the volatility and the interest rate parameters are assumed to be stochastic processes.


The  paper is structured as follows: in Section~\ref{sec_cppi_gme} we recall the key concepts related to CPPI and OBPI strategies and we introduce the new CPPI--GMEE approach in more details; in Section~\ref{sub:hist_sim} we compare the CPPI versus the CPPI with guaranteed minimum equity exposure approaches, mainly exploiting historical simulations; in Section~\ref{num_sim} we provide numerical experiments to show the differences of plain vanilla options, options on CPPI and options on CPPI--GMEE; in Section~\ref{concl} we recap results obtained in the paper also giving an outlook to related future developments.

\section{OBPI-CPPI hybrid portfolio allocation strategy} \label{sec_cppi_gme} 
Our idea is to introduce a principal protection strategy, which is a mixture of an OBPI and a CPPI approach. 
As a result of the merge between two portfolio insurance strategies, our proposal must first consider that the investor can not 
suffer losses related to the amount of her initial wealth. For this reason, in the following we will extensively make use of the 
so called \emph{protection level} (PL), defined as the percentage of the initial capital guaranteed at maturity. 

Besides, we will take into account the following key points: 

\begin{itemize}
	\item we invest a significant portion of the portfolio in time-congruent zero coupon bonds following a classical OBPI approach. 
				This assures to achieve the \emph{capital protection} at maturity; 
				
  \item the remaining part of the portfolio is put into an exotic call option linked to a CPPI strategy, where the CPPI portfolio 
				has an equity index as risky asset. This provides the \emph{participation in the equity market;} 
								
	\item we adjust the CPPI algorithm as underlying of the option in such a way that at any time the equity exposure will not fall 
				below a predefined level. This helps to avoid the occurrence of the well known \emph{cash-in risk.}
\end{itemize}

			%
			%

Before going into details about the concrete realization of the above described method, let us first describe the {\it standard} OBPI 
and CPPI portfolio allocation mechanisms. 
Throughout the paper, we let $0<T < \infty$ be the time horizon of the investment, while $\left(\Omega,\mathcal{F}, \FF, \PP \right)$ is a filtered probability space, with $\FF = \left\{\mathcal{F}_t\right\}_{0 \leq t \leq T},$ and we assume that all the processes  introduced in what follows are $\FF$-adapted. 

We consider a market consisting of a risky asset $S_t,$ whose dynamics will be specified later on, and a risk-less asset $B_t$ 
such that 
$$\de B_t = r_t B_t \de t \;.$$ 

\subsection{The OBPI portfolio allocation strategy} \label{def:obpi_strategy} 
The OBPI strategy is a portfolio insurance procedure characterized by ensuring a minimum terminal portfolio value, see, e.g., 
\cite{ZagKr11}. 

According with standard literature, see, e.g., \cite{BerPrig05},
we define the OBPI portfolio process $V^{OBPI} = \{\obpi\}_{t \,\in\, [0,\,T]},$ with initial value $\obpiz,$ as follows
$$\obpi = q B_t + p \, Call(t,S_t,K), \, \mbox{ for all } t\,\in\,[0,T] \;,$$ 
where $q$ represents the number of riskless assets acquired by the investor to protect the capital, $Call(t,S_t,K)$ is the 
call option at time $t$, written on $S_t,$, having strike price $K$ and maturity $T,$, while $p \geq 0$ is the number of calls which can be purchased 
at time $t = 0,$ given the risk budget, see, e.g., \cite{ASW13}. 

The OBPI approach is said to be \emph{static} in the sense 
that no trading occurs in $(0,T),$ so that  the unique portfolio values we are interested in are 
\begin{equation*}
\obpiz = q B_0 + p \, Call(0,S_0,K) \;\mbox{ and } \; \obpit = q K + p\max\left\{S_T - K, 0\right\} \;,
\end{equation*} 
therefore, at maturity, the client gets the capital $q K$ plus $p$ times any positive performance of $S_T$ greater than $K.$ 
In case $q = 1, \, p=1$ and $S_T > K,$ the client gets exactly the performance of the underlying asset.  


\subsection{The CPPI portfolio allocation strategy}\label{def:cppi_strategy}

In order to define the CPPI portfolio process we begin by 
specifying the so called \emph{floor} $F$ representing the lowest acceptable value of the portfolio. 
In particular, we consider the process $F = \left\{ F_t\right\}_{t\,\in\,[0,T]}$ with dynamic
$$\de F_t = r_t F_t \de t \;,$$ 
and initial value $F_0 = \mathcal{C} \exp\left\{\int_0^T r_u \de u \right\},$ where $\mathcal{C}$ is the fixed amount of 
capital guaranteed at maturity.  

We define the process $V^{CPPI} = \{\cppi\}_{t \,\in\, [0,\,T]}$ with initial value $\cppiz,$ representing the portfolio value 
associated to the CPPI strategy, namely 
\begin{equation} \label{cppi_ptf} 
\cppi = \alpha_t S_t + \beta_t B_t \;,
\end{equation}
where $\alpha_t,$ resp. $\beta_t,$ represents the portfolio proportion invested in the risky, resp. in the riskless asset. 

By assuming that the portfolio strategy is self-financing, the dynamics of the CPPI portfolio can be easily 
obtained from eq. \eqref{cppi_ptf} as follows 
\begin{equation} \label{cppi_ptf_2}
\de \cppi = \alpha_t \de S_t + \beta_t \de B_t \;. 
\end{equation} 
Moreover, we assume $\mathcal{C} < \cppiz \exp\left\{\int_0^T r_u \de u \right\},$ that is, the guaranteed return must be less than 
the market interest rate. 

Since we are interested in determining the optimal allocation, then,  
for all $t \,\in\, [0,\,T]$, we evluate the excess of the portfolio value $V^{CPPI}$ over the floor $F,$ dubbed \emph{cushion,} as 
\begin{equation} \label{cushion} 
C_t := 
\begin{cases}
\cppi - F_t, & \cppi \geq F_t \\ 
0, & \mbox{ otherwise }
\end{cases}
\;,
\end{equation}
so that $C_t = \max \{0; \cppi - F_t \}, \, \mbox{ for all } t\,\in\,[0,T].$  

The investment in stock represents the \emph{exposure}, which is given by 
\begin{equation} \label{def:exposure}
E_t = M \cdot C_t = M \cdot \max\left\{\cppi - F_t; 0 \right\}, \, \mbox{ for all } t\,\in\,[0,T] \,, 
\end{equation}
the constant $M$ being a \emph{multiplier} representing  the factor by which the risk budget is amplified, giving rise to the 
risky asset. 

\begin{remark}
Let us note that since we are dealing with a
dynamic leverage adjustment mechanism, if we consider a general setting, the multiplier can be represented in terms of a suitable continuous function $M_t,$ depending on different model parameter, see, e.g., \cite{Schied14}. While, for the sake of simplicity, in our case we will consider a constant multiplier $M := \frac{1}{ONR},$ 
where ONR factor represents the \emph{Over-Night risk} of the risky asset. 
The market practice usually assumes $ONR = 25\%$ for a given equity index that serves as underlying, implying that $M = 4.$ 
\end{remark}




\subsection{The CPPI with Guaranteed Minimum Equity Exposure} \label{sub:cashin}

As regards the evaluation of the percentage of wealth to be invested in the risky asset, 
eq. \eqref{def:exposure} gives 
\begin{align}\label{alpha_t}
\alpha_t = \frac{M \cdot C_t}{V_t}, \quad\quad \beta_t = \cppi - \alpha_t \;, 
\end{align}
by assuming $\cppi \neq 0$ $\PP$-a.e. 

In particular, eq. \eqref{alpha_t} implies that the investment in the risky asset might be potentially unbounded. 
To limit such a potential leverage effect in the optimal allocation, the market practice suggests to introduce 
the so called \emph{maximum leverage factor} $L_{max}$ in the equity weights $\alpha_t,$ such that  
\begin{equation} \label{alpha_eq}
\alpha_t := \max \left\{\min\left\{L_{max}; \frac{M \cdot C_{t}}{V_{t}^{CPPI}} \right\}; 0\right\} \;. 
\end{equation} 
Motivated by some regulatory constraints, see, e.g., \cite{UCITS_IV} 
, $L_{max}$ is tipically 
setted to $L_{max}=150\%$, or $L_{max}=200\%.$ 

However, it may happen that, due to sudden events, the price of the risky asset, on which the value of the CPPI portfolio depends, is significantly reduced. 
As a consequence, the value of the portfolio will be lower than the bound assigned through the floor, and from that moment on the manager will be able to allocate wealth only in the risk-less security until the contract expires. This situation is referred to as \emph{cash-in risk.} 

To avoid such a scenario, we introduce the so called \emph{guaranteed minimum equity exposure} 
$\alpha_{min}$ with $0 \leq \alpha_{min} \leq 1$ in eq. \eqref{alpha_eq}, obtaining
\begin{equation} \label{cppi_gme}
\alpha_t^{CPPI} = \max \left\{\min \left\{ L_{max}; \frac{M \cdot C_{t}}{V_{t}^{CPPI}} \right\}; \alpha_{min} \right\} \;.
\end{equation}
Consequently, the CPPI portfolio whose equity exposure is given by eq. \eqref{cppi_gme}, is labeled as \emph{CPPI--GMEE} portfolio. 

\begin{remark} 
Let us note that exploiting equations \eqref{alpha_eq} and \eqref{cppi_gme}, we easily get  that the case $\alpha_{min} = 0$ coincides with the traditional 
CPPI logic. 
\end{remark}

\section{Historical simulation of structured products with CPPI vs. CPPI with guaranteed minimum equity exposure} \label{sub:hist_sim}
In order to verify weather the presence of the GMEE threshold permits to dodge cash-in events, 
in this section we simulate CPPI portfolio allocation strategies, with and without guaranteed minimum equity exposure, linked to 
European equity markets. To capture the sensitivities of the proposed strategies, we propose two different scenarios. In the first case we consider $PL = 100\%$, in the time 
interval 2007 to 2017, whereas, in the second example, we analyze the $PL = 90\%$ case, referring to the time period 2000 to 2010. 

\paragraph{The 100\%-protection case.} The CPPI simulation has been conducted assuming the following 
\begin{itemize}
\item the underlying is given by a \emph{Euroland large and mid cap equity index} as risky asset between 31st of December 2007 
			and 29th of December 2017; 

\item the CPPI mechanism should protect $100\%$ of the initial investment after 10 years. For the sake of simplicity, we assume a 
			constant risk-free rate of $3\%,$ since the risk-free rates in 2008 were close to such a level. 
			The latter would have also determined the risk budget of the associated strategy. Moreover, starting with a new 
			CPPI within today's financial scenario, hence taking into consideration interest rates' levels near to zero, the related risk 
			budget would be also close to zero. In this framework, one can either extend the investment time horizon, to benefit from 
			higher interest rates deriving from considering related longer maturities, or set up the investment w.r.t. a lower protection 
			level, e.g. at $90\%,$ instead of $100\%;$
			
\item according to the CPPI methodology, we choose a multiplier $M=3$ and a maximum leverage factor of $L_{max}=150\%$, while 
			the guaranteed minimum equity exposure $\alpha_t$ is set to $30\%.$
\end{itemize}

\begin{figure}[t!]
\begin{center}
\includegraphics[width=12cm]{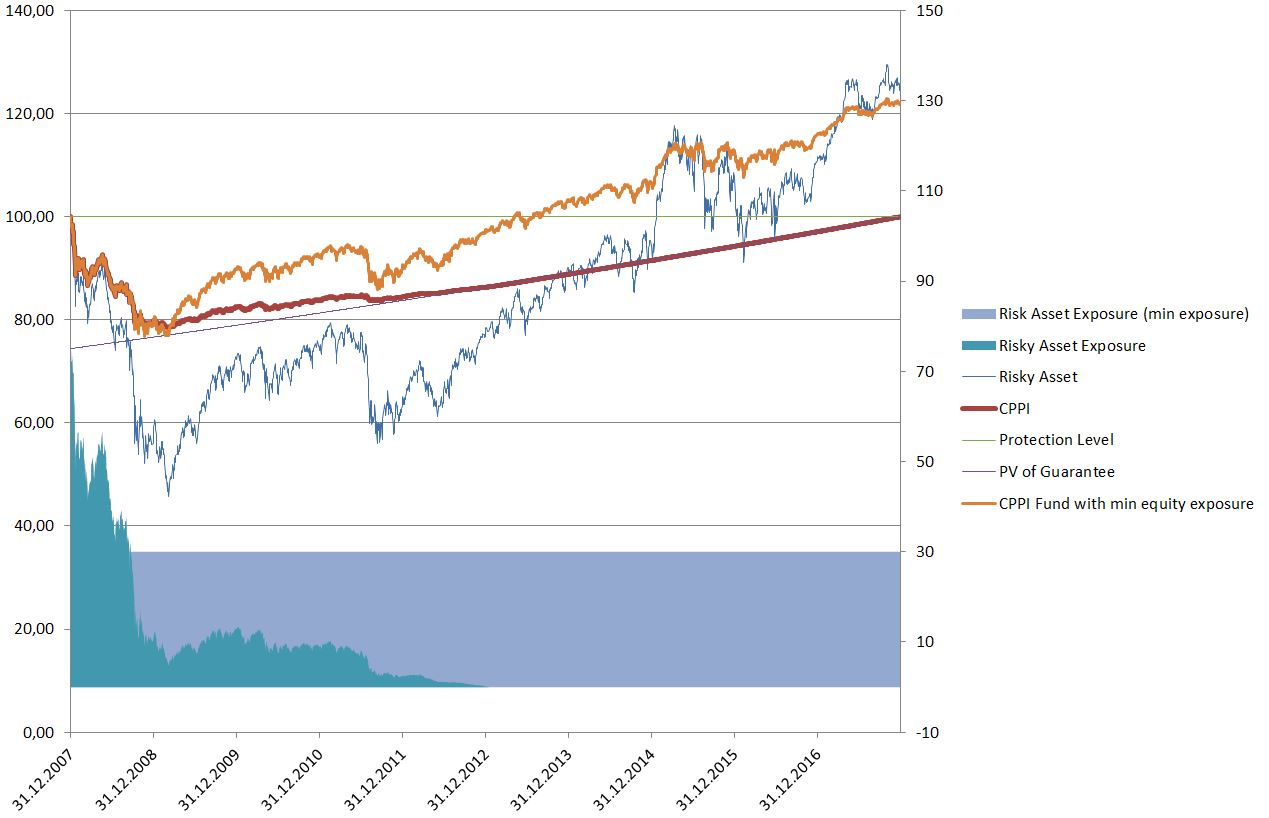}
\caption{Historic Simulation for a standard CPPI and a CPPI with guaranteed minimum equity exposure linked to European equity 
markets between 2007 and 2017.}\label{fig:Simulation_Case1} 
\end{center}
\end{figure}

In Fig.~\ref{fig:Simulation_Case1} we report the obtained results to better motivate our key idea. 

The left hand axis gives the performance of the risky asset, of a standard CPPI approach and a CPPI approach with guaranteed 
minimum equity exposure, also providing the present value of the guarantee in percentage of the initial investment. 
The right hand axis shows the risky asset exposure over time for a CPPI and a CPPI--GMEE portfolio in percentage of the 
overall portfolio allocation. 

The key findings can be summarized as follows. Looking at the risky asset itself, we see that the considered market index 
significantly lost in value between 2007 and March 2009. In fact, during this period the index lost close to $50\%$ of its 
initial value. After that, the equity index recovered nicely and over the full 10 year horizon the index generated a positive 
performance of more than $20\%.$ Nevertheless, from the point of view of a conservative investor, an equity index might 
be too volatile. 
Focusing on a traditional CPPI allocation logic, we can see that the red line gives the performance of a CPPI strategy linked 
to this equity index as risky asset, exploiting the parameters mentioned above. For the sake of simplicity, we do not consider 
transaction costs. 

The standard CPPI has an initial exposure to the risky asset of more than $70\%.$ Due to the extreme losses in the risky asset, 
the risk budget quickly decreases, and the CPPI needs to reduce the risky asset exposure to less than $30\%$ after 10 months. 
The CPPI approach itself can limit the losses successfully compared to the pure risky asset investment in this time, but it cannot 
participate in any upwards markets afterward. After some further volatilities in the risky asset over the next 4 years, we can see 
that the risky asset allocation drops to zero at the end of 2012, hence no market participation exists afterward. 
Overall, the CPPI can achieve $PL = 100\%,$ but it cannot benefit from the overall positive return in the equity market. 
The CPPI with guaranteed minimum equity exposure starts with the same risky asset allocation as the standard CPPI. 
Also this CPPI approach is forced to reduce its risky asset exposure significantly due to negative risky asset performance. 
Nevertheless, by definition, the risky asset exposure never drops below the predefined threshold of $30\%.$ 
Therefore, the proposed CPPI alternative approach can benefit from rising equity markets again and over the full remaining life time, 
the CPPI with guaranteed minimum equity exposure achieves a return which is quite comparable to the pure risky asset. 

\paragraph{The 90\%-protection case.} For the CPPI has been conducted assuming the following 
\begin{itemize}
\item to hold a Euroland large and mid cap equity index as risky asset, considered  between the 31st of December 1999 and 
			the 31st of December 2009;

\item the considered CPPI mechanism should protect $90\%$ of the initial investment after 10 years, and, for the sake of clarity, we 
			assume a constant risk-free rate of $4\%;$

\item within the CPPI we choose conservative multiplier $M=2$, capping the maximum leverage factor at $L_{max}=100\%$. 
			The guaranteed minimum equity exposure $\alpha_t$ is set at $30\%,$ and we do not consider any transaction costs. 
\end{itemize}

\begin{figure}[t!]
\begin{center}
\includegraphics[width=12cm]{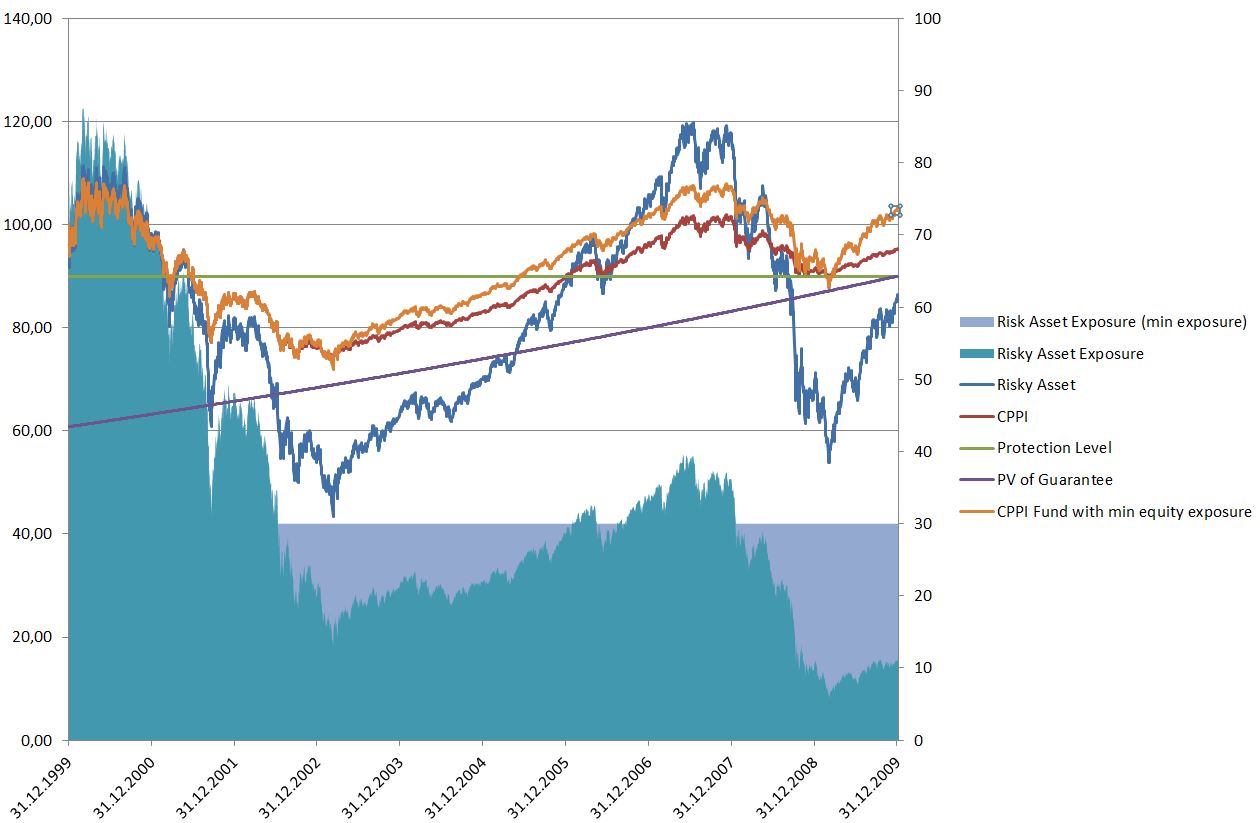}
\caption{Historic Simulation for a standard CPPI and a CPPI with guaranteed minimum equity exposure linked to European equity markets between 2000 and 2010.} \label{fig:Simulation_Case2}
\end{center}
\end{figure}

As can be seen in Fig.~\ref{fig:Simulation_Case2},  
the risky asset during this period of time starts positively, then achivieng a return of close to $20\%$, in the first year. 
But, between 2000 and 2002, the equity index loses about $50\%$ of  its initial value. 
Then, it well recovers until mid of 2007, when the worldwide financial crisis causes new severe losses. 
Therefore, after 10 years, the index loses about $15\%$ of its initial value. 

The standard CPPI linked to this index has an initial exposure of about  $80\%,$ as a result of higher risk budget given 
the lower protection level of just $90\%$ and a higher risk-free rate of $4\%$ compared to the previous case. 
When the equity index increases initially in value then the exposure of the standard CPPI increases to close to $90\%.$ 
But, during following years, the exposure is significantly reduced to less than $20\%$ in 2003. 
When markets are recovering, also the risky asset exposure increases to roughly $40\%,$ but during the financial crisis it 
falls below $10\%$ in early 2009. Therefore, after 10 years, the exposure is at $10\%$ and the performance of the CPPI ends 
with $-5\%,$ namely it could achieve a higher return than the index itself and than the aspired capital protection level of 
$90\%,$ but still the investor made a loss. 

The CPPI with guaranteed minimum equity exposure shows initially a similar behavior like the traditional CPPI. 
It also starts with an equity exposure of around $80\%,$ which increases to even $90\%$ and then falls to the guaranteed
 minimum equity exposure of $30\%$ in 2002. The equity exposure remains there, between 2005 and 2007 it increases again 
to ca. $40\%.$ 
During the financial crisis it drops back again, but, by definition, not below the minimum exposure of $30\%.$ 
As a result, the CPPI with guaranteed minimum equity exposure could limit losses from equity markets in 2001 and 2002, 
and also during the financial crisis. Nevertheless, it could especially benefit from recovering markets more than a standard CPPI.
After 10 years the new CPPI approach generated a positive return of more than $3\%.$ 
Moreover, it could achieve $PL = 90\%$ and it could generate a much better risk-return profile than 
a pure equity investment.

\section{Numerical Pricing of Options on CPPI with guaranteed minimum equity exposure} \label{num_sim}
By eq. \eqref{cppi_gme} we have that, even in case of a severe market drop, the equity participation will never go below 
$\alpha_{min}$, while, at the same time it would mean that this adjusted CPPI allocation implemented in a real portfolio 
might not be able to protect the invested capital. 

This justifies the second step of our proposal, namely our CPPI mechanism is always applied in an OBPI-based portfolio approach, 
in which the call option is linked to the CPPI allocation logic with guaranteed minimum equity exposure. 
Thus, we now consider options on CPPI and CPPI--GMEE strategies. More precisely, in what follows we provide the mathematical 
setting of our options on CPPI-GMEE as well as in depth numerical analysis related to the its implementation in the option 
pricing context. 

\subsection{The financial model} \label{model_dyn}
Assuming there exists a measure $\QQ \sim \PP,$ with respect to the filtered probability space $\left(\Omega, \mathcal{F}, \FF, \QQ \right)$ consider a portfolio allocation strategy in which a financial agent invest his wealth in one riskless asset, e.g. 
a bond, and in one risky asset, e.g., an equity index, over a time interval  $[0,\,T].$ More precisely,
we take a risk-free asset $B_t$ whose dynamics reads as follows
\begin{equation} \label{rf_asset}
\de B_t  = r_t B_t \de t \;,
\end{equation}
with a return $r_t$ at time $t,$ and a risky asset $S_t,$ such that 
\begin{align} \label{model}
\de S_t & = r_t S_t \de t + \sqrt{v_t} S_t \de Z^{S}_t \;, \\ 
\de v_t & = k (\theta - v_t) \de t + \sigma_v \sqrt{v_t} \de Z^{v}_t \;, \\ 
\de r_t & = \nu (\beta - r_t) \de t + \sigma_r v_t^{\gamma} \de Z^{r}_t \;.
\end{align}
The stochastic processes $Z^{S}_t, \; Z^{v}_t, \; Z^{r}_t $ are three correlated $(\FF,\QQ)$-adapted Wiener processes, with 
\begin{align*}
corr\left(\de Z^{S}_t, \de Z^{v}_t \right) & = \rho_{S,v} \;,\\ 
corr\left(\de Z^{S}_t, \de Z^{r}_t \right) & = \rho_{S,r} \;, \\ 
corr\left(\de Z^{v}_t, \de Z^{r}_t \right) & = \rho_{v,r} = \rho_{S,v}\rho_{S,r} \;,
\end{align*}
where $v_t$, resp. $r_t,$ represents the volatility, resp. the interest rate, stochastic process, with positive speed of 
reversion $k$, resp. $\nu$, long-term mean levels $\theta>0$, resp. $ \beta > 0$, and variance $\sigma_v$, resp. $\sigma_r$.

In such an economy, we consider a European contingent claim $X$ with maturity $T > 0,$ whose payoff is a real-valued function 
$f = f(S_t,v_t,r_t).$ The payoff function $f$ may either depend just on the final value of the underlyings, namely $S_T$ and 
$v_T,$ or relies upon the whole underlying path over $[0,T].$ In the former case we refer to plain vanilla 
call/put options, otherwise we might consider path-dependent options, e.g., considering options on CPPI. 

At any rate, we can define the price of the contingent claim as 
\begin{equation} \label{cont_claim} 
\OO_t(X) = \frac{1}{B_t} \media_t^{\QQ} \left[f(S_T,v_T,r_T) \right] \,,\, \forall t \,\in\, [0, T]\,,
\end{equation}
where $\media_t^{\QQ}$ is the conditional expectation taken w.r.t. the initial filtration $\FF$ to which the Wiener processes 
$Z^{S}_t, \, Z^{r}_t, \, Z^{v}_t$ have been defined to be adapted and under the risk-neutral measure $\QQ \sim \PP.$ 

Equivalently, we may consider a contingent claim $\hat{X}$ with maturity $T > 0,$ whose payoff is a real-valued function 
$f = f(V_t^{CPPI},v_t,r_t),$ relying upon the CPPI portfolio strategy. More precisely, we assume that the underlying asset 
for $\hat{X}$ is measured in units of the CPPI strategy, instead of units of stock, see, e.g., \cite{AlbStebWall17,EKZ11} 
for further details. 
In this case, 
the price of the contingent claim $\hat{X}$, at time $t$ reads as follows 
\begin{equation} \label{cont_claim_cppi} 
\hat{\OO}_t(X) = \frac{1}{B_t} \media_t^{\QQ} \left[f(V^{CPPI}_T,v_T,r_T) \right] \,,\, \forall t \,\in\, [0,T]\;. 
\end{equation}

\subsection{Options on CPPI and CPPI-GMEE} \label{option_pricing}
In this section we are going to compare At-The-Money (ATM) European call/put option prices, evaluated both in the standard case, 
i.e. when the underlying is the stock dynamics, and when the underlying equals the CPPI portfolio allocation strategy, 
the second type of computations has been conducted both with and without Guaranteed Minimum Equity Exposure. 
Let us assume the following: 
\begin{itemize}
	\item for normal CPPI-based strategies, the $PL = 100\%$ at the end of the option life time, the multiplier 
				is $M = 4$, while the Maximum Leverage is equal to $L_{max} = 150\%;$ 
				
	\item for options linked to CPPI strategies with guaranteed minimum equity exposure, the protection level, the multiplier and 
				the Maximum Exposure are as in the previous case, while the guaranteed minimum equity exposure is  $\alpha_{min} = 30\%.$ 
\end{itemize}
We assume no transaction costs and dividends are directly reinvested into the strategy. We also assume that all the CPPI strategies 
are re-allocated each business day. From a practical perspective, this implies that the manager assumes that trading actions are 
discretized in time, namely according to a $[0,T]$ subdivision of the following type  $0 = t_0 < t_1 < \ldots < t_n = T$, 
where $t_i$, $i=0, \ldots,n$ represent fixed trading dates, so that we have
\begin{equation} \label{cppi_ptf_disc}
V^{CPPI}_{t_i} = \alpha_{t_i} V	^{CPPI}_{t_{i-1}} \frac{S_{t_i} - S_{t_{i-1}}}{S_{t_{i-1}}} + (1 - \alpha_{t_i}) V^{CPPI}_{t_{i-1}} 
(1+r_{t_i})  \;,
\end{equation}
with initial condition $V^{CPPI}_0 = V_0.$ 

\begin{remark} 
There does not exist a fixed rule to determine the time grid used for rebalancing the CPPI portfolio. 
In the market practice often a so called \emph{trading filter} is applied, namely, as long the real allocation deviates 
not significantly to the theoretical CPPI allocation, no trading happens. This is to avoid trading on noise and also to reduce 
transaction costs. 

Overall a weekly allocation is probably a reasonable assumption. In case of financial distress the rebalancing frequency can 
increase, and managers might have to trade daily or, in extreme cases, even two or three times a day. 
\end{remark}

Concerning the numerical values for  the parameters, we consider four time horizons (measured in years) 
$T = \{1, \, 2, \, 5, \, 10 \},$ an initial interest rate $r_0 = 5\%,$ and an initial volatility level $v_0 = 20\%.$ 
The remaining parameters are described in Table~\ref{table:parameters}. 
\begin{table}[t!]
\centering
\caption{Model parameters used in the numerical experiments. The second column refers to the stochastic interest rate model 
(Vasicek model). The last column refers to the stochastic volatility model (Heston model). } \label{table:parameters}
\begin{tabular}{ccc}
\hline 
& Vasicek model & Heston model \\ 
\hline
Long-run mean & $\beta = 0.05$ & $\theta = 0.04$ \\ 
Rate of mean reversion & $k = 1.25$ & $\nu = 1.25$ \\ 
Volatility & $\sigma_r = 0.025$ & $\sigma_v = 0.2$ \\ 
Correlation & $\rho_{S,r} = -0.2$ & $\rho_{S,v} = -0.5$ \\ 
\hline
\end{tabular}
\end{table}

\subsubsection{Comparison of call options}\label{ATM_call}
\begin{figure}[h!]
\centering
\includegraphics[scale=0.78]{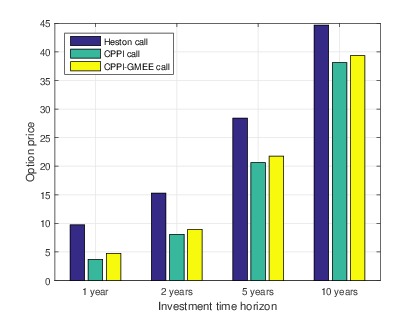} 
\caption{ATM call option pricing under different underlyings and different maturities. The Heston model parameters are: 
$\sigma_0 = 0.2, \, k = 1.25, \, \theta = \sigma_0^2, \, \sigma_v = 0.2, \, \rho_{S,v} = -0.5.$ The interest rate model 
parameters are: $r_0 = 0.05, \, \nu = 1.25, \, \beta = r_0, \, \sigma_r = 0.025, \, \gamma = 0.5, \, \rho_{S,r} = -0.2.$ 
The CPPI strategy parameters are: $L_{max} = 100\%, \, \alpha_{min} = 30\%, \, M = 4, \; PL = 100\%.$ } \label{comparison_C}
\end{figure}

In what follows, numerical results for the CPPI strategy exploited as European call option underlying are provided.
To better understand sensitivities, we compare the values of a European call option when the underlying 
is a pure risky asset, within the Vasicek-Heston model, the CPPI strategy, and the CPPI with guaranteed minimum equity exposure, 
respectively. 
In Fig.~\ref{comparison_C} we report the call option price for different maturities in each of the aforementioned scenarios. 

Considering an initial volatility level of $20\%$ and an initial interest rate level of $5\%,$ we observe that the price 
obtained by taking the pure risky asset as the derivative's underlying results is the most expensive strategy. 
Instead, the option pricing with respect to the CPPI strategy leads to a reduction of the option price. 
As expected, the CPPI with guaranteed minimum equity exposure leads to higher prices rather than the simple CPPI, 
this due to the {\it cost} we pay for having fixed a lower threshold of the risky asset exposure at $30\%.$  

In order to stress the role of the guaranteed minimum exposure, we study the call option value as a function of the
parameter $\alpha_{min}$ for different maturity levels. 

\begin{figure}[h!]
\centering
\includegraphics[scale=0.78]{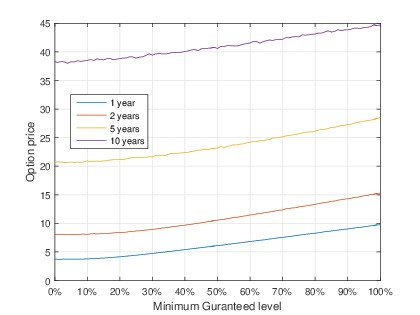} 
\caption{ATM call option price when the underlying is a CPPI strategy with guaranteed minimum equity exposure as a function 
of the minimum exposure parameter. The CPPI strategy parameters are: $L_{max} = 100\%, \, M = 4, \; PL = 100\%.$} \label{cppi_vs_gme_C}
\end{figure}

In Fig.~\ref{cppi_vs_gme_C} we observe that the option price increases as the minimum guarantee threshold raises. 

The lowest price is reached when the minimum guaranteed exposure is zero, which corresponds to the case of an usual CPPI strategy 
as underlying of the option. 
The case $\alpha_{min} = 100\%$ coincides with the plain vanilla call option, as in this case the allocation to risky asset is 
always 100\%. Since we assumed $L_{max} = 100\%,$ the effect of a rising option price with a rising minimum guaranteed equity 
exposure is also true for longer maturities. 

In Table~\ref{table_call} we report the results obtained by exploiting the
different allocation strategies for the evaluation of the call option prices, w.r.t. different initial interest rates and 
volatilities, and considering a short investment time horizon, i.e. taking $T=1$ year. 
In Panel A we reported the plain vanilla call option prices, while Panel B contains the normal CPPI-based options, 
and Panel C refers to the options linked to the CPPI with guaranteed minimum equity exposure. 
\begin{table}[t!]
\centering
\caption{ATM call option prices for different values of initial interest rate and initial volatility. 
The Heston model parameters are: 
$k = 1.25, \, \theta = \sigma_0^2, \, \sigma_v = 0.2, \, \rho_{S,v} = -0.5$, while the interest rate model parameters are: 
$\nu = 1.25, \, \beta = r_0, \, \sigma_r = 0.025, \, \gamma = 0.5, \, \rho_{S,r} = -0.2,$ and the CPPI strategy parameters are: $L_{max} = 100\%, \, \alpha_{min} = 30\%, \, M = 4, \; PL = 100\%.$} \label{table_call} 
\begin{adjustbox}{max width=\textwidth}
\begin{tabular}{cccccc}
\hline
\multicolumn{6}{c}{Panel A: option on pure risky asset } \\
\hline
\multicolumn{1}{c}{Initial interest rate $(r_0)$} & \multicolumn{5}{c}{Initial annual volatility $(v_0)$} \\ 
\hline
 & 0.10 & 0.20 & 0.30 & 0.40 & 0.50 \\
\hline
 0.01 & 5.29 &  9.19 & 13.06 & 16.89 & 20.68 \\
 0.03 & 6.07 &  9.77 & 13.57 & 17.38 & 21.18 \\
 0.05 & 6.84 & 10.35 & 14.14 & 17.88 & 21.57 \\
 0.07 & 7.65 & 11.06 & 14.73 & 18.38 & 22.01 \\
 0.10 & 8.93 & 12.05 & 15.61 & 19.24 & 22.88 \\
\hline 
\hline
\multicolumn{6}{c}{Panel B: option on CPPI } \\
\hline
\multicolumn{1}{c}{Initial interest rate $(r_0)$} & \multicolumn{5}{c}{Initial annual volatility $(v_0)$} \\ 
\hline
 & 0.10 & 0.20 & 0.30 & 0.40 & 0.50 \\
\hline
 0.01 & 2.64 & 2.64 & 2.65 & 2.65 & 2.62 \\
 0.03 & 3.72 & 3.72 & 3.71 & 3.72 & 3.66 \\
 0.05 & 4.78 & 4.78 & 4.78 & 4.78 & 4.72 \\
 0.07 & 5.82 & 5.82 & 5.84 & 5.82 & 5.86 \\
 0.10 & 7.35 & 7.35 & 7.35 & 7.35 & 7.42 \\
\hline
\hline
\multicolumn{6}{c}{Panel C: option on CPPI with Guaranteed Minimum Equity Exposure } \\
\hline
\multicolumn{1}{c}{Initial interest rate $(r_0)$} & \multicolumn{5}{c}{Initial annual volatility $(v_0)$} \\ 
\hline
 & 0.10 & 0.20 & 0.30 & 0.40 & 0.50 \\
\hline
 0.01 & 3.02 & 3.97 & 5.07 & 6.28 &  7.55 \\
 0.03 & 3.97 & 4.74 & 5.76 & 6.94 &  8.27 \\
 0.05 & 4.94 & 5.55 & 6.50 & 7.64 &  8.96 \\
 0.07 & 5.92 & 6.43 & 7.31 & 8.38 &  9.65 \\
 0.10 & 7.41 & 7.79 & 8.56 & 9.65 & 10.90 \\
\hline
\end{tabular}
\end{adjustbox}
\end{table}
We would like to underline that, while 
the call options linked to traditional CPPI-based approach remain almost constant for different volatility levels,
no matter about the initial interest rate value, higher volatilities might increase the option price for a pure risky 
asset underlying, and higher volatilities for a CPPI strategy increase the risk of a cash-in event such that a higher number 
of simulated paths ends up with the minimum protection level of $100\%.$ 
As a result, the option price is less depending on volatility levels. 

Let us note that options linked to CPPI-based strategies are significantly cheaper than plain vanilla ones, 
thanks to the embedded risk management features. Moreover, below an annual market volatility of $20\%$ the CPPI with 
and without minimum exposure give a comparable price range. When the volatility exceeds $20\%$ the former becomes more expensive. 

\subsubsection{Comparison of put options}
In this subsection we provide numerical results for the put option case.

\begin{figure}[h!]
\centering
\includegraphics[scale=0.78]{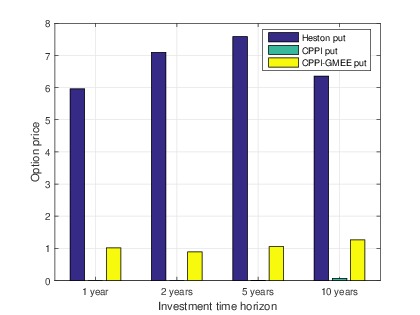}
\caption{ATM put option pricing under different underlyings and different maturities. The Heston model parameters are: 
$\sigma_0 = 0.2, \, k = 1.25, \, \theta = \sigma_0^2, \, \sigma_v = 0.2, \, \rho_{S,v} = -0.5,$ 
while the interest rate model parameters are: $r_0 = 0.05, \, \nu = 1.25, \, \beta = r_0, \, \sigma_r = 0.025, \, \gamma = 0.5, 
\, \rho_{S,r} = -0.2,$ and the CPPI strategy parameters are: $L_{max} = 100\%, \, \alpha_{min} = 30\%, \, M = 4, \; PL = 100\%.$ } 
\label{comparison_P}
\end{figure}

Fig.~\ref{comparison_P} shows the put option price for different maturities when the underlying is represented by a pure 
risky asset, resp. by a standard CPPI portfolio, resp. by a CPPI strategy with guaranteed minimum equity exposure. 

\begin{figure}[h!]
\centering
\includegraphics[scale=0.78]{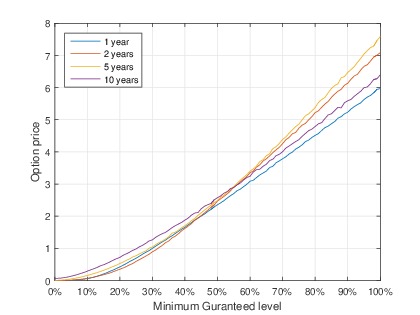} 
\caption{ATM put option price when the underlying is a CPPI strategy with guaranteed minimum equity exposure as a function 
of the minimum exposure parameter. The CPPI strategy parameters are: $L_{max} = 100\%, \, \alpha_{min} = 30\%, \, M = 4, 
\; PL = 100\%.$} \label{cppi_vs_gme_P}
\end{figure}

As expected, the put option over the standard CPPI strategy provides a zero-value price. 
Interestingly, we see some cases in which also the value of the put option linked to a CPPI is positive, i.e. 
we have some paths for which the CPPI logic does not achieve the protection level of $100\%.$
The latter is due to the fact that, especially for longer time horizon, the risk increases that in certain cases 
the overnight loss is higher than the assumptions embedded in the multiplier. 
As seen in Sect.~\ref{ATM_call}, we obtain that the put options linked to CPPI-based strategies are cheaper than 
the standard derivative options based on a pure risky underlying. 

Moreover, see Fig.~\ref{cppi_vs_gme_P}, we study the put option value as a function of the minimum exposure $\alpha_{min}$ 
for different maturity levels. As before, the case of $\alpha_{min} = 0\%$ coincides with the put on a standard CPPI. 
That is why the resulting option values are close to zero. For $\alpha_{min} = 100\%$ the results coincide with put option 
prices using a pure risky asset as underlying. 

Finally, in Table~\ref{table:put} we evaluate the put option prices for different initial interest rates and volatilities, 
for an investment time $T=1$ year. Panel A refers to put options on a pure risky asset, while Panel B refers to the 
standard CPPI case, and Panel C reports data for a CPPI with guaranteed minimum equity exposure. 
It can be seen that put options linked to CPPI-based strategies with minimum exposure to the risky asset are positive 
and significantly more expansive than put options on a normal CPPI. These results show, as expected, there exists a number of 
paths in which the CPPI with guaranteed minimum exposure can lead to real losses and cannot achieve capital preservation like 
the standard CPPI approach. 

\begin{table}[t!]
\centering
\caption{ATM put option prices for different values of initial interest rate and initial volatility. 
The Heston model parameters are: $k = 1.25, \, \theta = \sigma_0^2, \, \sigma_v = 0.2, \, \rho_{S,v} = -0.5.$ 
The interest rate model parameters are: $\nu = 1.25, \, \beta = r_0, \, \sigma_r = 0.025, \, \gamma = 0.5, \, 
\rho_{S,r} = -0.2.$ The CPPI strategy parameters are: $L_{max} = 100\%, \, \alpha_{min} = 30\%, \, M = 4, \; 
PL = 100\%.$} \label{table:put} 
\begin{adjustbox}{max width=\textwidth}
\begin{tabular}{cccccc}
\hline
\multicolumn{6}{c}{Panel A: option on pure risky asset } \\
\hline
\multicolumn{1}{c}{Initial interest rate $(r_0)$} & \multicolumn{5}{c}{Initial annual volatility $(v_0)$} \\ 
\hline
 & 0.10 & 0.20 & 0.30 & 0.40 & 0.50 \\
\hline
 0.01 & 2.62 & 6.49 & 10.35 & 14.18 & 18.02 \\
 0.03 & 2.29 & 5.99 &  9.80 & 13.57 & 17.38 \\
 0.05 & 1.97 & 5.49 &  9.25 & 13.00 & 16.74 \\
 0.07 & 1.69 & 5.09 &  8.78 & 12.43 & 16.12 \\
 0.10 & 1.36 & 4.49 &  8.06 & 11.68 & 15.33 \\

\hline 
\hline
\multicolumn{6}{c}{Panel B: option on CPPI } \\
\hline
\multicolumn{1}{c}{Initial interest rate $(r_0)$} & \multicolumn{5}{c}{Initial annual volatility $(v_0)$} \\ 
\hline
 & 0.10 & 0.20 & 0.30 & 0.40 & 0.50 \\
\hline
0.01 & 0.00 & 0.00 & 0.00 & 0.00 & 0.00 \\
0.03 & 0.00 & 0.00 & 0.00 & 0.00 & 0.00 \\
0.05 & 0.00 & 0.00 & 0.00 & 0.00 & 0.00 \\
0.07 & 0.00 & 0.00 & 0.00 & 0.00 & 0.02 \\
0.10 & 0.00 & 0.00 & 0.00 & 0.00 & 0.07 \\
\hline
\hline
\multicolumn{6}{c}{Panel C: option on CPPI with Guaranteed Minimum Equity Exposure } \\
\hline
\multicolumn{1}{c}{Initial interest rate $(r_0)$} & \multicolumn{5}{c}{Initial annual volatility $(v_0)$} \\ 
\hline
 0.01 & 0.37 & 1.31 & 2.41 & 3.62 & 4.93 \\
 0.03 & 0.24 & 1.01 & 2.03 & 3.21 & 4.51 \\
 0.05 & 0.15 & 0.76 & 1.71 & 2.85 & 4.13 \\
 0.07 & 0.09 & 0.59 & 1.47 & 2.56 & 3.84 \\
 0.10 & 0.05 & 0.42 & 1.20 & 2.27 & 3.53 \\
\hline
\end{tabular}
\end{adjustbox}
\end{table}

\subsubsection{CPPI-based option pricing strategies for different protection levels}
\begin{figure}[t!]
\centering
{\includegraphics[width=0.95\linewidth]{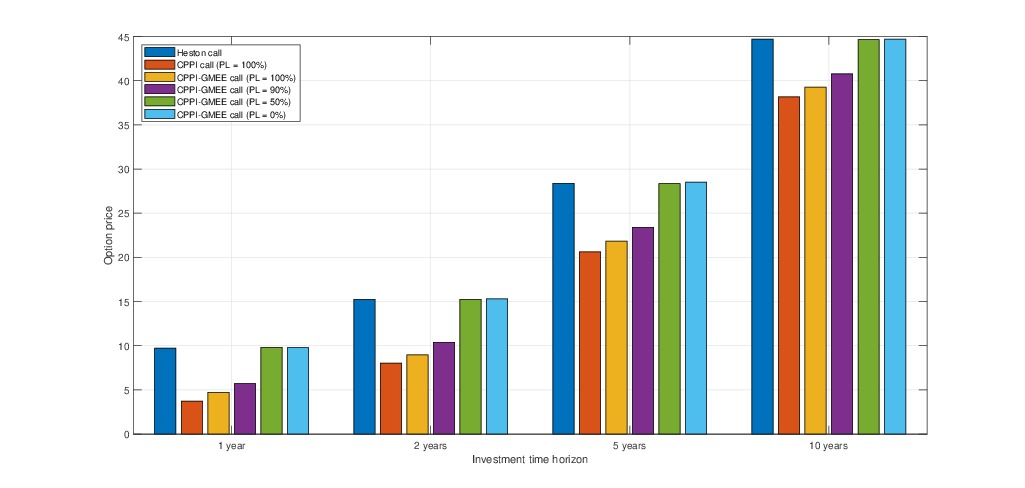}
}
\caption{ATM call option pricing under different underlyings, different maturities and different protection levels. 
The Heston model parameters are: $\sigma_0 = 0.2, \, k = 1.25, \, \theta = \sigma_0^2, \, \sigma_v = 0.2, \, \rho_{S,v} = -0.5,$ 
while the interest rate model parameters are: $r_0 = 0.05, \, \nu = 1.25, \, \beta = r_0, \, \sigma_r = 0.025, \, \gamma = 0.5, 
\, \rho_{S,r} = -0.2,$ and the CPPI strategy parameters are: $L_{max} = 100\%, \, \alpha_{min} = 30\%, \, M = 4, 
\; PL = [0 \%, \, 50\%, \, 90\%, \,100\%].$} \label{option_PV_call}
\end{figure}

\begin{figure}[t!]
\centering
{\includegraphics[width=0.95\linewidth]{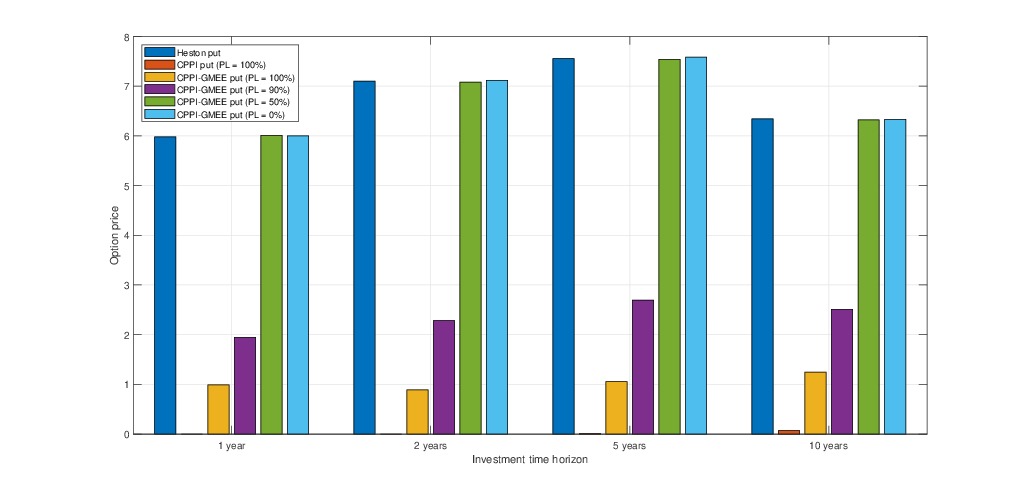}
}
\caption{ATM put option pricing under different underlyings, different maturities and different protection levels. 
The Heston model parameters are: $\sigma_0 = 0.2, \, k = 1.25, \, \theta = \sigma_0^2, \, \sigma_v = 0.2, \, \rho_{S,v} = -0.5,$ 
while  the interest rate model parameters are: $r_0 = 0.05, \, \nu = 1.25, \, \beta = r_0, \, \sigma_r = 0.025, \, \gamma = 0.5, 
\, \rho_{S,r} = -0.2,$ and the CPPI strategy parameters are: $L_{max} = 100\%, \, \alpha_{min} = 30\%, \, M = 4, 
\; PL = [0 \%, \, 50\%, \, 90\%, \,100\%].$} \label{option_PV_put}
\end{figure}

An alternative way to adjust the portfolio allocation is to modify the protection level. 
More precisely, we consider a CPPI-GMEE allocation with different protection levels, ranging between $0\%$ and $100\%.$  
A reduction of the protection level increases the risk budget and, consequently, the equity exposure of the portfolio. 
For the standard CPPI strategy the protection level remains at $100\%.$
The numerical results for ATM call/put options are provided in  Fig.~\ref{option_PV_call} and ~\ref{option_PV_put}. 

  
We observe that: 
\begin{itemize}
	\item by reducing the protection level of the CPPI-GMEE approach from $100\%$ to $90\%$ the CPPI-GMEE strategy gets riskier. 
				This implies that the corresponding option price increases significantly. 
				The same behavior can be spotted when the protection level is even more reduced, e.g. when we consider $PL = 50\%.$ 
				
				Such a circumstance is more evident in the case of put options, where the option price doubles when the 
				protection level halves; 
	
	\item the case in which $PL = 0\%$ equals the case with a pure risky asset as the derivative underlying, hence we see 
				the square option price;
	
	\item there exists the risk that the CPPI strategy ends below $100,$ implying that also the put option price on standard 
				CPPI is greater than zero for long maturities. 
\end{itemize}


\section{Conclusion} \label{concl} 
In this paper we have introduced a novel extension within the set of
investment insurance strategies by combining an OBPI approach with a CPPI logic, reflecting a minimum guaranteed 
equity exposure.  
Indeed, besides using the CPPI portfolio as an underlying of suitable options to meet investors' capital protection needs, 
we consider the so called Guaranteed Minimum Equity Exposure, 
hence imposing that the percentage of wealth invested in the risky security cannot fall below a fixed threshold. 
This allows to provide a CPPI-based strategy avoiding the cash-in event after 
the critical situation where the price of the underlying suddenly collapses, because of, e.g., market shocks. 
This represents a concrete innovation in the literature related to the 
portfolio protection strategies based on OBPI or CPPI. 

We have provided historical simulations showing how the risk-return profile changes, according to the market environment, 
and describing the option prices' behaviors under different frameworks, namely, when the underlying is  pure risky asset, 
a CPPI strategy, or a CPPI--GMEE based one. Obtained results clearly illustrate that, depending on the parameters choice, 
our method provides a valuable compromise between a pure risky asset investment strategy and a traditional CPPI one. 
In fact, it ensures to avoid the aforementioned cash-in risk of a standard CPPI, although it is rather more expensive 
than the options on standard CPPI. 

We would like to underline that the present work represents a first step in our research agenda. 
Further contributions, that are subject of our on-going research, include more structured derivatives evaluated with 
respect to general stochastic volatility models, also including the presence of jumps. 
Concurrently, we plan to study the sensitivity of the CPPI-GMEE approach to the changes of market parameters, 
and to compare options on CPPI with options on other dynamic asset allocation strategies, such as  the \emph{VolTarget} 
ones, also allowing the CPPI-GMEE to have lock-in elements. Finally, we intend to examine the role played by 
transaction costs in the option valuation on CPPI-GMEE framework.

%

\end{document}